\pdfoutput=1

\documentclass[twocolumn,aps,pra,superscriptaddress]{revtex4}

\usepackage{times}
\usepackage{graphicx}
\usepackage{epsfig}
\usepackage{amsfonts}
\usepackage{amsmath}
\usepackage{amssymb}
\usepackage{color}
\usepackage{multirow}
\usepackage[colorlinks=true,citecolor=blue,urlcolor=blue]{hyperref}
\usepackage{float}

\newcommand{\ket}[1]{|#1\rangle}

\begin{document}



\title{Testing noncontextuality inequalities that are building blocks of quantum correlations}


\date{\today}


\author{Mauricio~Arias}
\author{Gustavo~Ca\~nas}
\author{Esteban~S.~G\'omez}
\author{Johanna~F.~Barra}
\affiliation{Departamento de F\'{\i}sica, Universidad de Concepci\'on, 160-C Concepci\'on, Chile}
\affiliation{Center for Optics and Photonics, Universidad de Concepci\'on, Concepci\'on, Chile}
\affiliation{MSI-Nucleus for Advanced Optics, Universidad de Concepci\'on, Concepci\'on, Chile}
\author{Guilherme~B.~Xavier}
\affiliation{Center for Optics and Photonics, Universidad de Concepci\'on, Concepci\'on, Chile}
\affiliation{MSI-Nucleus for Advanced Optics, Universidad de Concepci\'on, Concepci\'on, Chile}
\affiliation{Departamento de Ingenier\'{\i}a El\'ectrica, Universidad de Concepci\'on, 160-C Concepci\'on, Chile}
\author{Gustavo~Lima}
\affiliation{Departamento de F\'{\i}sica, Universidad de Concepci\'on, 160-C Concepci\'on, Chile}
\affiliation{Center for Optics and Photonics, Universidad de Concepci\'on, Concepci\'on, Chile}
\affiliation{MSI-Nucleus for Advanced Optics, Universidad de Concepci\'on, Concepci\'on, Chile}
\author{Vincenzo D'Ambrosio}
 \affiliation{Dipartimento di Fisica, Sapienza
 Universit\`{a} di Roma, I-00185 Roma, Italy}
\author{Flavio Baccari}
 \affiliation{Dipartimento di Fisica, Sapienza
 Universit\`{a} di Roma, I-00185 Roma, Italy}
\author{Fabio Sciarrino}
 \affiliation{Dipartimento di Fisica, Sapienza
 Universit\`{a} di Roma, I-00185 Roma, Italy}
\author{Ad\'{a}n~Cabello}
\affiliation{Departamento de F\'{\i}sica Aplicada II, Universidad de Sevilla, E-41012 Sevilla, Spain}


\begin{abstract}
 Measurement scenarios containing events with relations of exclusivity represented by pentagons, heptagons, nonagons, etc., or their complements are the only ones in which quantum probabilities cannot be described classically. Interestingly, quantum theory predicts that the maximum values for any of these graphs cannot be achieved in Bell inequality scenarios. With the exception of the pentagon, this prediction remained experimentally unexplored. Here we test the quantum maxima for the heptagon and the complement of the heptagon using three- and five-dimensional quantum states, respectively. In both cases, we adopt two different encodings: linear transverse momentum and orbital angular momentum of single photons. Our results exclude maximally noncontextual hidden-variable theories and are in good agreement with the maxima predicted by quantum theory.
\end{abstract}




\maketitle


\section{Introduction}
\label{Sec0}


Most of the fascinating aspects and applications of quantum theory (QT) come from the fact that quantum probabilities cannot be explained by classical probability theory \cite{Gleason57,Specker60,Bell66,Bell64,KS67}. This impossibility is made evident by the violation of Bell \cite{Bell64} and noncontextuality (NC) \cite{Cabello08} inequalities and is fundamental for the advantages of quantum computation \cite{HWVE14} and quantum secure communication \cite{Ekert91}. Interestingly, a recent result \cite{CSW10,CSW14,AFLS12} reveals the basic components of genuinely quantum probabilities.

Before presenting this result we need to recall some concepts. An event is a transformation produced by a measurement that changes the quantum state assigned to the system prior to the measurement into a new one. For example, consider a Bell inequality experiment on pairs of particles initially prepared in a certain quantum state. Two observers, Alice and Bob, perform independent measurements on their respective particles. In this scenario, an event can be characterized as follows: ``Given the initial quantum state $\psi$, Alice measures observable $A_0$ and obtains outcome $0$ and Bob measures $B_0$ and obtains outcome $0$.'' We may denote this event by $(A_0=0,B_0=0|\psi)$. Two events are exclusive when they cannot happen simultaneously. For example, the events $(A_0=0,B_0=0|\psi)$ and $(A_0=1,B_1=0|\psi)$ are exclusive because $A_0=0$ and $A_0=1$ cannot happen simultaneously. The exclusivity graph of a set of events is the graph in which events are represented by vertices and exclusive events by connecting edges. For example, the exclusivity graph of the events $(A_0=0,B_0=0|\psi)$, $(A_0=1,B_1=0|\psi)$, $(A_0=0,B_1=1|\psi)$, $(A_1=0,B_1=0|\psi)$, and $(A_1=1,B_0=1|\psi)$ is a pentagon. An induced subgraph is a subset of the vertices of a graph together with any edges whose endpoints are both in this subset. The complement of a graph $G$ is the graph with the same vertex set as $G$ and such that two vertices are adjacent if and only if they are not adjacent in $G$.

A result in Refs.~\cite{CSW10,CSW14,AFLS12} establishes that the only experiments that cannot be described by classical probability theory are those whose exclusivity graphs have, as induced subgraphs, odd cycles $C_n$, with $n \ge 5$ (i.e., pentagons, heptagons, nonagons, etc.), or their complements $\overline{C_n}$. In light of this result, it is important to investigate, both theoretically and experimentally, the sets of quantum probabilities represented by odd cycles and their complements. Key questions within this research program are the following: (i) What are the maximum values allowed by QT for the sum of the probabilities of events whose exclusivity graphs are odd cycles or their complements? (ii) In what experiments do these maxima occur? (iii) What physical principles enforce these maxima?

The answer to question (i) was provided in Refs.~\cite{CSW10,CSW14}. The answer to question (ii) was provided in Ref.~\cite{CDLP13}. These results are reviewed in Sec.~\ref{Sec1}. In addition, it has been proven recently that none of the quantum maxima for odd cycles or their complements can be achieved in Bell inequality scenarios \cite{R15}. This result, also reviewed in Sec.~\ref{Sec1}, underlines the importance to experimentally test the quantum predictions for the scenarios in which these maxima are achieved. The answer to question (iii) is yet unknown, but there is a principle that relates the maxima of $C_n$ and $\overline{C_n}$. This result is also reviewed in Sec.~\ref{Sec1}.

The aim of this paper is to present the results of four experiments: Two of them independently test the maximum quantum value for $C_7$ using two different photonic degrees of freedom and the other two independently test the maximum quantum value for $\overline{C_7}$. The two experimental platforms used, each located in a different laboratory, are described in Sec.~\ref{Sec2} and the experimental results are presented in Sec.~\ref{Sec3}. Using two different degrees of freedom allows us to check the validity of the quantum predictions independently of the system. Testing the maximum values for both $C_7$ and $\overline{C_7}$ also allows us to test the connection between them as mentioned before. In Sec.~\ref{Sec4} we present a summary.


\section{Theoretical overview}
\label{Sec1}


This section provides an overview of the theoretical results that motivate our experiments.


\subsection{Quantum maximum for the heptagon}


The quantum maximum for $C_5$ occurs in the Klyachko-Can-Binicio\u{g}lu-Shumovsky (KCBS) scenario \cite{KCBS08}, which has been tested in several experiments \cite{LLSLRWZ11,UZZWYDDK13,AACB13,KSSWHZJDYD12,MANCB14}. For odd cycles on seven or more vertices, the quantum maxima can be realized in the violation of NC inequalities extending the KCBS inequality. The inequality for $C_7$ [see Fig.~\ref{Fig1}(a)] was first considered by Bengtsson \cite{Bengtsson09} and the general form was introduced in Refs.~\cite{CSW10,CSW14,LSW11}. These NC inequalities were proven to be tight in Ref.~\cite{AQBTC13}. For $C_7$, the NC inequality and its maximum quantum violation are
\begin{equation}
\label{I1}
 S(C_7)=\sum_{j=1}^7 P(1,0|j, j \oplus 1) \stackrel{\mbox{\tiny{ NCHV}}}{\leq} 3 \stackrel{\mbox{\tiny{ QT}}}{\leq} 3.3177,
\end{equation}
where $P(1,0|j, j \oplus 1)$ is the joint probability of obtaining $1$ and $0$ for sharp compatible measurements $j$ and $j \oplus 1$, respectively, $\oplus$ denotes sum modulo $7$, and NCHV indicates noncontextual hidden-variable theories.

The quantum maximum of $S(C_7)$ can be achieved with a three-dimensional quantum system (qutrit) in the initial state $|\psi\rangle$ and with the measurements $j = |u_j\rangle \langle u_j|$, where
\begin{subequations}
\begin{equation}
 \label{u}
 |\psi\rangle = \left( \begin{array}{c}
 1\\
 0\\
 0\end{array}\right),\;\;\;\;
 |u_j\rangle = \left( \begin{array}{l}
 \cos \left(\phi_1\right)\\
 \sin \left(\phi_1\right) \cos \left(\phi_2\right)\\
 \sin \left(\phi_1\right) \sin \left(\phi_2\right)\end{array}\right),
\end{equation}
with
\begin{equation}
 \cos \left(\phi_1\right) = \sqrt{\frac{\cos \left(\frac{\pi}{7}\right)}{1+\cos \left(\frac{\pi}{7}\right)}},\;\;\;\;
 \phi_2 = \frac{6 \pi j}{7},
\end{equation}
\end{subequations}
and $j=1,\ldots,7$. This result can be found in Refs.~\cite{Bengtsson09,LSW11,CSW10,CSW14}.

In contrast, the quantum maximum of $S(C_7)$ in Bell inequality scenarios is only
\begin{equation}
\label{P1p}
 S_{\rm QLM\,max}(C_7) = 2+ \frac{3 \sqrt{3}}{4} \approx 3.2990
\end{equation}
and can be achieved in the two-party, three-setting, two-outcome scenario. This result, which extends a similar observation made for $C_5$ in Ref.~\cite{SBBC13}, is proven in Ref.~\cite{R15} using tools of Ref.~\cite{RDLTC14}.


\subsection{Quantum maximum for the complement of the heptagon}

\begin{figure}[tb]
 \centering
 \includegraphics[width=0.48 \textwidth]{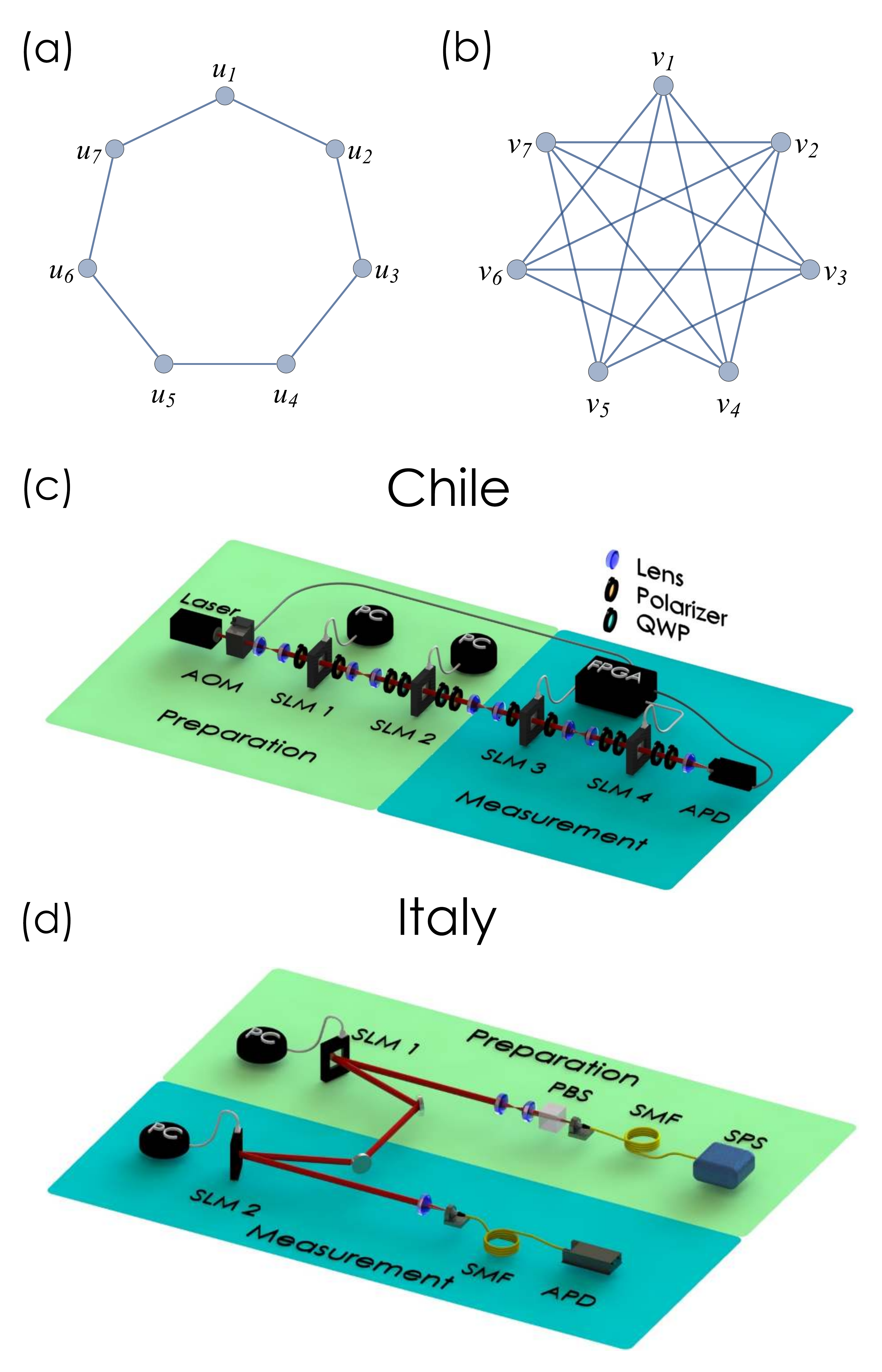}
 \caption{(Color online) (a) Cycle $C_7$ and (b) its complement $\overline{C_7}$. (c) Setup used in the experiments performed in Chile for testing $S(C_7)$ and $S(\overline{C_7})$ using linear transverse momentum of single photons. QWP stands for quarter wave plate, the rest of the notation is explained in the text. (d) Setup used in the experiments performed in Italy for testing $S(C_7)$ and $S(\overline{C_7})$ using orbital angular momentum states of single photons. \label{Fig1}}
\end{figure}

For the complements of odd cycles on seven or more vertices, the quantum maxima can be obtained in the maximum violations of the NC inequalities proposed in Ref.~\cite{CDLP13}. For $\overline{C_7}$ [see Fig.~\ref{Fig1}(b)], the NC inequality and its maximum quantum violation are
\begin{equation}
\label{I2}
 S(\overline{C_7})=\sum_{k=1}^7 P(1,0,0|k \ominus 2, k, k \oplus 2) \stackrel{\mbox{\tiny{ NCHV}}}{\leq} 2 \stackrel{\mbox{\tiny{ QT}}}{\leq} 2.1099,
\end{equation}
where $k \ominus 2$, $k$, and $k \oplus 2$ are sharp and compatible and $\oplus$ and $\ominus$ denote addition and subtraction modulo~7, respectively.

The quantum maximum of $S(\overline{C_7})$ can be achieved with a five-dimensional quantum system in the initial state $|\phi\rangle$ and with the measurements $k = |v_k\rangle \langle v_k|$, where
\begin{subequations}
\begin{equation}
\label{v}
 |\phi\rangle = \left( \begin{array}{c}
 1\\
 0\\
 0\\
 0\\
 0\end{array}\right),\;\;\;
 |v_k\rangle = \left( \begin{array}{l}
 \cos \left(\varphi_1\right)\\
 \sin \left(\varphi_1\right) \cos \left(\varphi_2\right) \cos \left(\varphi_3\right)\\
 \sin \left(\varphi_1\right) \cos \left(\varphi_2\right) \sin \left(\varphi_3\right)\\
 \sin \left(\varphi_1\right) \sin \left(\varphi_2\right) \cos \left(\varphi_4\right)\\
 \sin \left(\varphi_1\right) \sin \left(\varphi_2\right) \sin \left(\varphi_4\right)\\
 \end{array}\right),
\end{equation}
with
\begin{align}
 \cos \left(\varphi_1\right) &= \sqrt{\frac{1+\cos \left(\frac{\pi}{7}\right)}{7 \cos \left(\frac{\pi}{7}\right)}},\\
 \cos \left(\varphi_2\right) &= 2 \sin \left(\frac{\pi}{7}\right) \sqrt{\frac{2 \cos \left(\frac{\pi}{7}\right)}{-1+6 \cos \left(\frac{\pi}{7}\right)}},\\
 \varphi_3 &= \frac{2 \pi 2 k}{7},\;\;\;\;
 \varphi_4 = \frac{2 \pi k}{7},
\end{align}
\end{subequations}
and $k=1,\ldots,7$. This result was introduced in Ref.~\cite{CDLP13}.

Interestingly, this maximum cannot be achieved neither with quantum systems of smaller dimension \cite{CDLP13} nor within Bell inequality scenarios \cite{R15}. The precise value of the quantum maximum of $S(\overline{C_7})$ in Bell inequality scenarios is not yet known.


\subsection{Why these maxima?}


No physical principle has yet been able to explain the quantum maxima for $S(C_7)$ and $S(\overline{C_7})$. However, the exclusivity principle \cite{Cabello13,Yan13,ATC13,Cabello15} singles out the quantum maxima of the following NC inequality:
\begin{equation}
\begin{split}
\label{I3}
 &S(C_7 \otimes \overline{C_7}) \\
 &=\sum_{j=1}^7 \sum_{k=1}^7 P(1,0,1,0,0|j, j \oplus 1,k \ominus 2, k, k \oplus 2) \stackrel{\mbox{\tiny{ NCHV}}}{\leq} 6 \stackrel{\mbox{\tiny{ Q}}}{\leq} 7,
\end{split}
\end{equation}
where $\otimes$ denotes the OR graph product. The reason for the name $S(C_7 \otimes \overline{C_7})$ is that its exclusivity graph is $C_7 \otimes \overline{C_7}$. The vertex set of the graph $G_1 \otimes G_2$ is the Cartesian product $V(G_1) \times V(G_2)$, where $V(G_1)$ and $V(G_2)$ are the vertex sets of $G_1$ and $G_2$, respectively. Two vertices $(u_1, u_2)$ and $(v_1, v_2)$ of $G_1 \otimes G_2$ are connected by an edge if and only if the vertices $u_1$ and $v_1$ of $G_1$ are connected by an edge or if the vertices $u_2$ and $v_2$ of $G_2$ are connected by an edge. Interestingly, the quantum maximum for $S(C_7 \otimes \overline{C_7})$ can be achieved by combining in a single experiment one test targeting the quantum maximum for $S(C_7)$ on one physical system and another test targeting the quantum maxima for $S(\overline{C_7})$ on an independent system.


\section{Our two experimental setups}
\label{Sec2}


There are three types of experiments for testing NC inequalities \cite{GKCLKZGR10,LKGC11}. In experiments of type~1, compatibility is enforced by performing each measurement on a spatially separated system, as in Bell inequality experiments. However, as pointed out in the introduction, $S_{\rm max}(C_7)$ and $S_{\rm max}(\overline{C_7})$ cannot be observed in type~1 experiments. In type~2 experiments, sharp measurements are performed sequentially on the same system \cite{KZGKGCBR09}. Type~2 is the best option when type~1 is impossible. As shown below, targeting $S_{\rm max}(\overline{C_7})$ would require sequences of three sharp measurements on, at least, a five dimensional quantum system. Unfortunately, no existing experimental platform is yet mature enough for such an experiment. This is the reason why we adopted the type~3 approach as in Refs.~\cite{LLSLRWZ11,UZZWYDDK13,KSSWHZJDYD12,CEGSXLC14,CAEGCXL14}.

In type~3 experiments, measurements are not sequentially performed, but each context (i.e., set of compatible measurements) is implemented at once by means of a $d$-outcome measurement. The assumption is that each rank-one measurement corresponds to one outcome of the $d$-outcome measurement. However, this implies that each context requires a different $d$-outcome measurement, thus each rank-one measurement is implemented differently depending on the context. This is acceptable under the assumption that all implementations represent the same rank-1 measurement at the level of hidden variables.


\subsection{Linear transverse momentum setup}


For testing both $S(C_{7})$ and $S(\overline{C_7})$ we employ two different setups. The first setup exploits the linear transverse momentum (LTM) of single photons. A $d$-dimensional quantum state is created by defining $d$ path alternatives for the photon transmission through a diffractive aperture \cite{Neves05,Lima09}. Here $d=3$ for $S(C_{7})$ and $d=5$ for $S(\overline{C_7})$. To prepare state $\ket{\phi}$ we use a set of $d$ parallel slits dynamically generated using a sequence of spatial light modulators (SLMs). If the transverse coherence length of the beam is larger than the distance separating the slits, then the state of the transmitted photon is given by $\ket{\Psi} = \frac{1}{\sqrt{C}}\sum_{l=-l_d}^{l_d} \sqrt{t_{l}}e^{i\Phi_l}\ket{l}$, where $l_d=\frac{d-1}{2}$ and $\ket{l}$ represents the state of the photon transmitted by the \textit{l}th slit \cite{Neves05}. Here $t_l$ ($\Phi_l$) is the transmissivity (phase) defined for each slit and $C$ the normalization constant. The slits are $64$~$\mu$m wide and have a separation between them of $128$~$\mu$m. The advantage of using SLMs to define slits is that they allow us to control $t_l$ and $\Phi_l$ independently and dynamically for each slit.

The LTM experimental setup is depicted in Fig.~\ref{Fig1}(c). It has two stages: preparation and measurement. The preparation stage consists of an attenuated single-photon source (SPS) and of two SLMs used to encode $d$-dimensional quantum states. SLM~1 controls the real part of the coefficients of the prepared states, while SLM~2 controls their phases \cite{Lima10,Lima11,Lima13}. The measurement stage uses the same technique and employs two other SLMs (SLM~3 and SLM~4) to randomly project state $\ket{\phi}$ onto one of states $\ket{v_k}$. A field-programmable gate array electronics unit controls the entire experimental setup. For example, the probability $P_{\ket{\psi}}(1,0|j,j \oplus 1)$ is obtained as the ratio between the photon counts when state $\ket{\psi}$ is projected over state $\ket{u_j}$ and the total number of counts when state $\ket{\psi}$ is projected over the three-dimensional basis containing $\ket{u_j}$ and $\ket{u_{j \oplus 1}}$.


\subsection{Orbital angular momentum setup}


The second setup exploits the orbital angular momentum (OAM) of single photons. This degree of freedom, related to the spatial distribution of the electromagnetic field, provides a wide alphabet for qudits \cite{Moli07}. Orbital angular momentum eigenstates are indeed spatial modes with azimuthal phase factor $e^{il \phi}$, exhibiting a helical wavefront with $l$ intertwined helices of handedness given by the sign of $l$. Such a degree of freedom can provide qudits of arbitrary dimension encoded in single-photon states and has been so far exploited in several fields ranging from quantum information \cite{Damb12,Naga12,dimwit} to microscopy \cite{micro}, astrophysics \cite{Tamb}, metrology \cite{Damb13}, and optical tweezing \cite{tweez}.

The OAM experimental setup for testing $S(C_7)$ and $S(\overline{C_7})$ using OAM states is shown in Fig.~\ref{Fig1}(d). The logical basis is chosen as $\{\ket{-1},\ket{0},\ket{1}\}$ for $S(C_{7})$ and $\{\ket{-2},\ket{-1},\ket{0},\ket{1},\ket{2}\}$ for $S(\overline{C_7})$, where $\ket{l}$ identifies the state of a photon with OAM $l\hbar$. Photonic OAM states are manipulated with SLMs by means of holograms generated by computers that allow us to introduce an arbitrary phase retardation pattern in the beam. The preparation and measurement of arbitrary superposition OAM states requires us to tailor both phase and amplitude profiles of the beam. This double manipulation can also be achieved by acting only on the phase of the beam by inducing selective losses \cite{slm}. Here we exploit a recently developed holographic method specifically optimized to obtain high state fidelities \cite{qusix,BBSKB13}. Measurements are chosen so as to maximize the number of OAM eigenstates in the set in order to increase the overall diffraction efficiency of the holograms.

The setup in Fig.~\ref{Fig1}(d) has two stages. In the preparation stage, an SPS generates heralded single photons by spontaneous parametric down conversion in a beta-barium-borate crystal. Single photons are then projected on the fundamental TEM$_{00}$ mode by means of a single mode fiber (SMF). A polarizing beam splitter (PBS) selects the horizontally polarized photons, which are then prepared in the state $\ket{\psi}$ by using an SLM (SLM~1). In the measurement stage, projective measurements are performed by means of a second SLM (SLM~2), whose hologram is chosen in order to convert state $\ket{v_i}$ into TEM$_{00}$, which is the only one that can be coupled to an SMF. As a result, the combination of SLM~2, an SMF, and a single-photon detector, an avalanche photodiode (APD), allows us to perform the desired projection. For example, $P_{\ket{\phi}}(1,0,0|k \ominus 2,k,k \oplus 2)$ is obtained as the ratio between the photon counts when state $\ket{\phi}$ is projected over state $\ket{v_k}$ and the total number of counts when state $\ket{\phi}$ is projected over a five-dimensional basis containing $\ket{v_{k \ominus 2}}$, $\ket{v_{k}}$, and $\ket{v_{k \oplus 2}}$. An imaging system, not shown in Fig.~\ref{Fig1}(d), was implemented between the screens of the two SLMs in order to avoid the Gouy phase-shift effect.


\begin{figure}[tb]
 \centering
 \includegraphics[width=0.45 \textwidth]{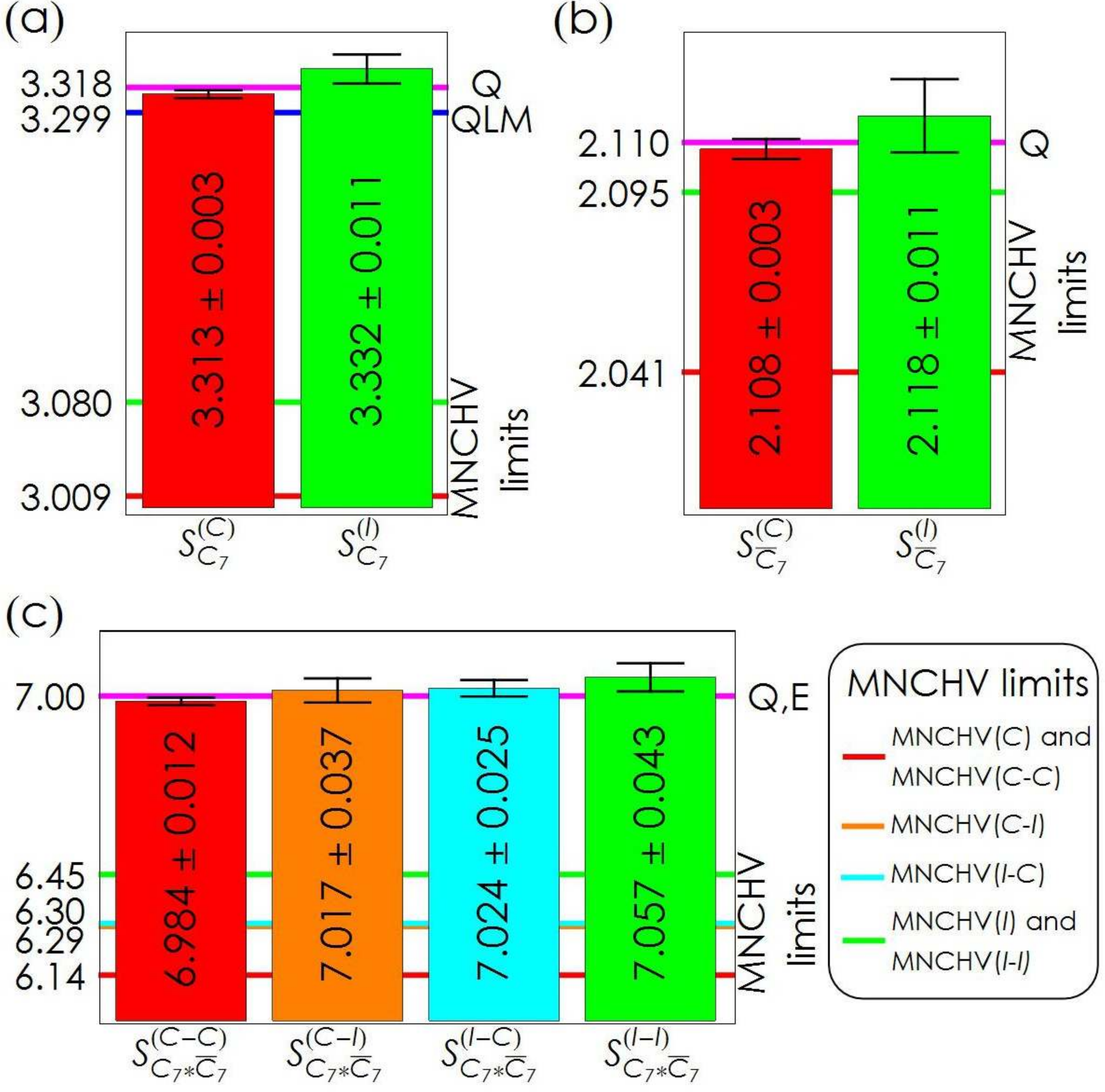}
 \caption{(Color online) (a) Experimental results of the tests of $S(C_7)$ performed in Chile (left) and Italy (right).
 (b) Experimental results of the tests of $S(\overline{C_7})$ performed in Chile (left) and Italy (right). Q denotes the maximum predicted by QT for an ideal experiment. QLM denotes the maximum using quantum local measurements on composite systems for an ideal experiment. This bound is only known for the case of $S(C_7)$. MNCHV indicates the upper bound for maximally noncontextual hidden-variable theories. This upper bound depends on the experimental errors and, therefore, is different for each experiment. Lines of different colors are used to indicate the MNCHV bound of each experiment. (c) Experimental results of the four combinations of experiments testing $S(C_7)$ and $S(\overline{C_7})$ used to test $S(C_7 \otimes \overline{C_7})$. E is the upper bound for theories satisfying the exclusivity principle.
 \label{Fig2}}
\end{figure}


\begin{table}[tb]
\centering
\begin{tabular}{ccc}
 \hline \hline
 $(j,j \oplus 1)$ & $P_{\ket{\psi}}(1,0|j,j \oplus 1)$ & Theory \\
 \hline
 $(1,2)$ & $0.488\pm 0.003$ & $0.474$ \\
 $(2,3)$ & $0.455\pm 0.003$ & $0.474$ \\
 $(3,4)$ & $0.486\pm 0.003$ & $0.474$ \\
 $(4,5)$ & $0.467\pm 0.003$ & $0.474$ \\
 $(5,6)$ & $0.478\pm 0.003$ & $0.474$ \\
 $(6,7)$ & $0.476\pm 0.003$ & $0.474$ \\
 $(7,1)$ & $0.462\pm 0.003$ & $0.474$ \\
 \hline
 $S(C_7)$ & $3.313 \pm 0.003$ & $3.318$ \\
 \hline \hline
\end{tabular}
\caption{\label{Table1a} Experimental results for $S(C_7)$ using LTM in Chile. Errors have been calculated assuming Poissonian statistics. Column 3 indicates the results predicted by quantum theory for an ideal experiment.}
\end{table}


\begin{table}[tb]
\centering
\begin{tabular}{ccc}
 \hline \hline
 $(j,j \oplus 1)$ & $P_{\ket{\psi}}(1,0|j,j \oplus 1)$ & Theory \\
 \hline
 $(1,2)$ & $0.462\pm 0.007$ & $0.474$ \\
 $(2,3)$ & $0.479\pm 0.007$ & $0.474$ \\
 $(3,4)$ & $0.458\pm 0.008$ & $0.474$ \\
 $(4,5)$ & $0.482\pm 0.008$ & $0.474$ \\
 $(5,6)$ & $0.449\pm 0.011$ & $0.474$ \\
 $(6,7)$ & $0.488\pm 0.011$ & $0.474$ \\
 $(7,1)$ & $0.513\pm 0.008$ & $0.474$ \\
 \hline
 $S(C_7)$ & $3.332 \pm 0.011$ & $3.318$ \\
 \hline \hline
\end{tabular}
\caption{\label{Table1b} Experimental results for $S(C_7)$ using OAM in Italy.}
\end{table}


\begin{table}[tb]
\centering
\begin{tabular}{ccc}
 \hline \hline
 $(k \ominus 2,k,k \oplus 2)$ & $P_{\ket{\phi}}(1,0,0|k \ominus 2,k,k \oplus 2)$ & Theory \\
 \hline
 $(1,3,5)$ & $0.296\pm 0.001$ & $0.301$ \\
 $(2,4,6)$ & $0.306\pm 0.001$ & $0.301$ \\
 $(3,5,7)$ & $0.308\pm 0.001$ & $0.301$ \\
 $(4,6,1)$ & $0.295\pm 0.001$ & $0.301$ \\
 $(5,7,2)$ & $0.304\pm 0.001$ & $0.301$ \\
 $(6,1,3)$ & $0.309\pm 0.001$ & $0.301$ \\
 $(7,2,4)$ & $0.291\pm 0.001$ & $0.301$ \\
 \hline
 $S(\overline{C_7})$ & $2.108 \pm 0.003$ & $2.110$ \\
 \hline \hline
\end{tabular}
\caption{\label{Table2a} Experimental results for $S(\overline{C_7})$ using LTM in Chile.}
\end{table}


\begin{table}[tb]
\centering
\begin{tabular}{ccc}
 \hline \hline
 $(k \ominus 2,k,k \oplus 2)$ & $P_{\ket{\phi}}(1,0,0|k \ominus 2,k,k \oplus 2)$ & Theory \\
 \hline
 $(1,3,5)$ & $0.317\pm 0.006$ & $0.301$ \\
 $(2,4,6)$ & $0.330\pm 0.010$ & $0.301$ \\
 $(3,5,7)$ & $0.315\pm 0.006$ & $0.301$ \\
 $(4,6,1)$ & $0.281\pm 0.010$ & $0.301$ \\
 $(5,7,2)$ & $0.261\pm 0.006$ & $0.301$ \\
 $(6,1,3)$ & $0.286\pm 0.009$ & $0.301$ \\
 $(7,2,4)$ & $0.327\pm 0.009$ & $0.301$ \\
 \hline
 $S(\overline{C_7})$ & $2.118 \pm 0.011$ & $2.110$ \\
 \hline \hline
\end{tabular}
\caption{\label{Table2b} Experimental results for $S(\overline{C_7})$ using OAM in Italy.}
\end{table}


\section{Experimental results and discussion}
\label{Sec3}


The experimental results for $S(C_7)$ and $S(\overline{C_7})$ are summarized in Figs.~\ref{Fig2}(a) and \ref{Fig2}(b), respectively. The experimental probabilities leading to these results are detailed in Tables~\ref{Table1a}--\ref{Table2b}. The experimental results for $S(C_7 \otimes \overline{C_7})$, obtained by combining the four possible ways the experiments in Chile and Italy are summarized in Fig.~\ref{Fig2}(c).

Inequalities (\ref{I1}), (\ref{I2}), and (\ref{I3}) are derived under the assumption that observables are compatible which implies, e.g. for the experiment testing $S(C_7)$, that $P_{\ket{\psi}}(1,\_|j,j \oplus 1)=P_{\ket{\psi}}(\_,1|j \ominus 1,j)$, where $P_{\ket{\psi}}(1,\_|j,j \oplus 1)$ is the marginal probability of obtaining result 1 for $j$ when $j$ is measured together with $j \oplus 1$ (i.e., in the context $\{j,j \oplus 1\}$). However, due to experimental imperfections, this condition is not exactly satisfied in the experiment. Therefore, a more detailed analysis of the experimental results is needed before reaching any conclusion.

Recent works have investigated what the experimental violation of an NC inequality actually proves when the experimental imperfections are taken into account \cite{Winter14,DKL15}. Here, to answer the question of what our experimental violations of inequalities (\ref{I1}), (\ref{I2}), and (\ref{I3}) prove, we have followed the approach in Ref.~\cite{DKL15}. This approach has the advantage of allowing us to reach conclusions about the impossibility of hidden-variable theories {\em without assuming QT} (see Ref.~\cite{DKL15} for a discussion of the advantages of the approach in Ref.~\cite{DKL15} with respect to the one in Ref.~\cite{Winter14}).

In Ref.~\cite{DKL15} it is shown that experimental results of a violation of NC inequalities can reveal the impossibility of explaining these results with maximally noncontextual hidden-variable (MNCHV) theories. This occurs if the experimental results violate a new inequality containing the same sum of probabilities of events, but in which the noncontextual bound is increased by an amount $\epsilon > 0$, which depends on the experimental results. The analysis introduced in Ref.~\cite{DKL15}, applied to inequalities (\ref{I1}), (\ref{I2}), and (\ref{I3}), is as follows.

The bound for $S(C_7)$ for NCHV theories in inequality (\ref{I1}) is valid assuming that the (hidden) value of measurement $j$ is the same regardless of whether $j$ is measured with $j \oplus 1$ or with $j \ominus 1$. That is,
$P(j_{\{j,j \oplus 1\}} \neq j_{\{j \ominus 1,j\}}) = 0$. However, these probabilities are experimentally unaccessible. Fortunately, they are lower bounded by an experimentally testable quantity since
\begin{equation}
 P(j_{\{j,j \oplus 1\}} \geq j_{\{j \ominus 1,j\}}) \ge T(j_{\{j,j \oplus 1\}},j_{\{j \ominus 1,j\}}),
 \label{eq2}
\end{equation}
where
\begin{equation}
 T(j_{\{j,j \oplus 1\}},j_{\{j \ominus 1,j\}}) = \frac{1}{2}\sum_x |P(x,\_|j,j \oplus 1)-P(\_,x|j-1,j)|.
\end{equation}
If, as it is our case, $x \in \{0,1\}$, then $P(1,\_|j,j \oplus 1) = 1 - P(0,\_|j,j \oplus 1)$ and, therefore, $T(j_{\{j \ominus 1,j\}},j_{\{j,j \oplus 1\}})=|P(\_,1|j \ominus 1,j)-P(1,\_|j,j \oplus 1)|$.

Similarly, the bound for $S(\overline{C_7})$ for NCHV theories in inequality (\ref{I2}) is valid assuming that the (hidden) value of measurement $k$ is the same regardless of whether $k$ is measured with $k \ominus 4$ and $k \ominus 2$, with $k \ominus 2$ and $k \oplus 2$, or with $k \oplus 2$ and $k \oplus 4$ (in this experiment $k$ is in three contexts). In this case, e.g., $T(k_{\{k \ominus 4,k \ominus 2,k\}},k_{\{k \ominus 2,k,k \oplus 2\}})=|P(\_,\_,1|k \ominus 4,k \ominus 2,k)-P(\_,1,\_|k \ominus 2,k,k \oplus 2)|$ and
\begin{eqnarray}
P(k_{\{k \ominus 4,k \ominus 2,k\}},
k_{\{k \ominus 2,k,k \oplus 2\}},
k_{\{k,k \oplus 2,k \oplus 4\}}\;\text{not all equal}) \nonumber \\
\geq \frac{1}{2}[
T(k_{\{k \ominus 4,k \ominus 2,k\}},k_{\{k \ominus 2,k,k \oplus 2\}}) \nonumber \\
+ T(k_{\{k \ominus 4,k \ominus 2,k\}},k_{\{k,k \oplus 2,k \oplus 4\}}) \nonumber \\
+ T(k_{\{k \ominus 2,k,k \oplus 2\}},k_{\{k,k \oplus 2,k \oplus 4\}})]. \label{eq3}
\end{eqnarray}

Following Ref.~\cite{DKL15}, we define maximally noncontextual hidden-variable (MNCHV) theories as those in which equalities in Eqs.~(\ref{eq2}) and (\ref{eq3}) hold. Then we can test whether or not the experimental results can be described by MNCHV theories simply by checking whether or not they satisfy the following inequalities:
\begin{subequations}
\begin{equation}
S(C_7) \stackrel{\mbox{\tiny{ MNCHV}}}{\leq} 3 + \epsilon(C_7),
\end{equation}
\begin{equation}
S(\overline{C_7}) \stackrel{\mbox{\tiny{ MNCHV}}}{\leq} 2 + \epsilon(\overline{C_7}),
\end{equation}
\end{subequations}
where,
\begin{subequations}
\begin{equation}
 \epsilon(C_7) = \frac{1}{2} \sum_{j=1}^7 T(j_{\{j \ominus 1,j\}},j_{\{j,j \oplus 1\}}),
\end{equation}
\begin{eqnarray}
 \epsilon(\overline{C_7}) = \frac{1}{2}\sum_{k=1}^7 [T(k_{\{k \ominus 4,k \ominus 2,k\}},k_{\{k \ominus 2,k,k \oplus 2\}}) \nonumber \\ + T(k_{\{k \ominus 4,k \ominus 2,k\}},k_{\{k,k \oplus 2,k \oplus 4\}})
\nonumber \\ + T(k_{\{k \ominus 2,k,k \oplus 2\}},k_{\{k,k \oplus 2,k \oplus 4\}})].
\end{eqnarray}
\end{subequations}

In the experiments performed in Chile, $\epsilon_{\rm exp}(C_7)=0.0089$ and $\epsilon_{\rm exp}(\overline{C_7})=0.041$. In the experiments performed in Italy, $\epsilon_{\rm exp}(C_7)=0.08$ and $\epsilon_{\rm exp}(\overline{C_7})=0.095$. Here $\epsilon_{\rm exp}(C_7 \otimes \overline{C_7})$ is estimated under the assumption that experiments testing $S(C_7)$ and the experiment testing $S(\overline{C_7})$ are statistically independent, as is the case in the tests combining the results of different experiments (some of them even on different physical systems in different laboratories).

The conclusions for the experiments testing $S(C_7)$ using LTM in Chile and OAM in Italy are summarized in Fig.~\ref{Fig2}(a). The conclusions for the experiments testing $S(\overline{C_7})$ using LTM in Chile and OAM in Italy are summarized in Fig.~\ref{Fig2}(b). The conclusions for the four experiments testing $S(C_7 \otimes \overline{C_7})$ are summarized in Fig.~\ref{Fig2}(c). In all the cases, the experimental results exclude MNCHV theories. The results of the two experiments testing $S(C_7)$ cannot be explained with local quantum measurements and therefore certify values that cannot be obtained in Bell inequality scenarios, thus confirming this fundamental prediction of QT. All results are in good agreement with the maxima for $S(C_7)$, $S(\overline{C_7})$, and $S(C_7 \otimes \overline{C_7})$ predicted by QT and, therefore, with the maximum for $S(C_7 \otimes \overline{C_7})$ allowed by the exclusivity principle.


\section{Conclusion}
\label{Sec4}


We have tested the maximum quantum values of two types of quantum correlations that have been identified as basic components of the quantum correlations that cannot be explained classically. Interestingly, according to QT, neither of these two maxima can be achieved with local measurements on composite systems, i.e., in Bell inequality experiments; they can only occur in specific NC inequality experiments. Here we have observed them by testing two different NC inequalities. Each experiment was performed on two different physical systems and using two different setups in order to verify the universality of the quantum predictions. Our results, analyzed with state-of-the-art theoretical tools, confirm the predictions of QT.

In addition, in order to investigate the physical principles that may be behind these quantum limits, we have tested a third NC inequality obtained by considering the two systems as a single composite system. The quantum maximum for this inequality equals the maximum allowed by any theory satisfying the exclusivity principle. Our results have also confirmed that this value is experimentally reachable.

We hope that our experiments will stimulate further developments to observe fundamental quantum correlations under conditions that avoid some of the assumptions made in this work. Specifically, it would be interesting to test quantum correlations in experiments with sequential quantum projective measurements on indivisible quantum systems of high dimensionality. Another interesting line of research would be experimentally certifying nonlocality of measurements by exploiting gaps between the quantum maxima with local measurements and the corresponding quantum maxima.


\begin{acknowledgments}
We thank J.-\AA.\ Larsson, R.\ Rabelo, M.\ Terra Cunha, and A.\ Winter for useful conversations. This work was supported by FONDECYT Grant No.\ 1120067, Milenio Grant No.\ RC130001, PIA-CONICYT Grant No.\ PFB0824 (Chile), Projects No.\ FIS2011-29400 and ``Advanced Quantum Information'' (MINECO, Spain) with FEDER funds, the FQXi large grant project ``The Nature of Information in Sequential Quantum Measurements,'' the program Science without Borders (CAPES and CNPq, Brazil), the Starting Grant 3D-QUEST through Grant No.\ 307783 (www.3dquest.eu). M.\ A., G.\ C., E.\ S.\ G., and J.\ F.\ B. acknowledge financial support from CONICYT.
\end{acknowledgments}




\begin{thebibliography}{99}


\bibitem{Gleason57}
 A. M. Gleason,
 J. Math. Mech. \textbf{6}, 885 (1957).

\bibitem{Specker60}
 E. P. Specker,
 \href{http://dx.doi.org/10.1111/j.1746-8361.1960.tb00422.x}{Dialectica \textbf{14}, 239 (1960);}
 [English translation: \href{http://arxiv.org/abs/1103.4537}{\eprint{arXiv:1103.4537}].}

\bibitem{Bell66}
 J. S. Bell,
 \href{http://dx.doi.org/10.1103/RevModPhys.38.447}{Rev. Mod. Phys. \textbf{38}, 447 (1966).}

\bibitem{Bell64}
 J. S. Bell,
 Physics (Long Island City, NY) \textbf{1}, 195 (1964).

\bibitem{KS67}
 S. Kochen and E. P. Specker,
 J. Math. Mech. \textbf{17}, 59 (1967).


\bibitem{Cabello08}
 A. Cabello,
 \href{http://dx.doi.org/10.1103/PhysRevLett.101.210401}{Phys. Rev. Lett. \textbf{101}, 210401 (2008).}


\bibitem{HWVE14}
 M. Howard, J. Wallman, V. Veitch, and J. Emerson,
 \href{http://dx.doi.org/10.1038/nature13460}{Nature (London) \textbf{510}, 351 (2014).}

\bibitem{Ekert91}
 A. K. Ekert,
 \href{http://dx.doi.org/10.1103/PhysRevLett.67.661}{Phys. Rev. Lett. \textbf{67}, 661 (1991).}


\bibitem{CSW10}
 A. Cabello, S. Severini, and A. Winter,
 \href{http://arxiv.org/abs/1010.2163}{\eprint{arXiv:1010.2163}.}

\bibitem{CSW14}
 A. Cabello, S. Severini, and A. Winter,
 \href{http://dx.doi.org/10.1103/PhysRevLett.112.040401}{Phys. Rev. Lett. \textbf{112}, 040401 (2014).}

\bibitem{AFLS12}
 A. Ac\'{\i}n, T. Fritz, A. Leverrier, and A. B. Sainz,
 \href{http://dx.doi.org/10.1007/s00220-014-2260-1}{Comm. Math. Phys. \textbf{334}, 533 (2015).}


\bibitem{CDLP13}
 A. Cabello, L. E. Danielsen, A. J. L\'opez-Tarrida, and J. R. Portillo,
 \href{http://dx.doi.org/10.1103/PhysRevA.88.032104}{Phys. Rev. A \textbf{88}, 032104 (2013).}


\bibitem{R15}
 R. Rabelo, M. Terra Cunha, and A. Cabello
 (in preparation).


\bibitem{KCBS08}
 A. A. Klyachko, M. A. Can, S. Binicio\u{g}lu, and A. S. Shumovsky,
 \href{http://dx.doi.org/10.1103/PhysRevLett.101.020403}{Phys. Rev. Lett. \textbf{101}, 020403 (2008).}


\bibitem{LLSLRWZ11}
 R. {\L}apkiewicz, P. Li, C. Schaeff, N. Langford, S.~Ramelow, M.~Wie\'sniak, and A.~Zeilinger,
 \href{http://dx.doi.org/10.1038/nature10119}{Nature (London) \textbf{474}, 490 (2011).}

\bibitem{UZZWYDDK13}
 M. Um, X. Zhang, J. Zhang, Y. Wang, S. Yangchao, D.-L. Deng, L.-M. Duan, and K. Kim,
 \href{http://dx.doi.org/10.1038/srep01627}{Sci. Rep. \textbf{3}, 1627 (2013).}

\bibitem{AACB13}
 J. Ahrens,
 E. Amselem,
 A. Cabello, and
 M. Bourennane,
 \href{http://dx.doi.org/10.1038/srep02170}{Sci. Rep. \textbf{3}, 2170 (2013).}

\bibitem{KSSWHZJDYD12}
 X. Kong, M. Shi, F. Shi, P. Wang, P. Huang, Q. Zhang, C. Ju, C. Duan, S. Yu, and J. Du,
 \href{http://arxiv.org/abs/1210.0961}{\eprint{arXiv:1210.0961}.}

\bibitem{MANCB14}
 B. Marques, J. Ahrens, M. Nawareg, A. Cabello, and M. Bourennane,
 \href{http://dx.doi.org/10.1103/PhysRevLett.113.250403}{Phys. Rev. Lett. \textbf{113}, 250403 (2014).}


\bibitem{Bengtsson09}
 I. Bengtsson,
 in
 {\em Foundations of Probability and Physics-5}, edited by
 L. Accardi, G. Adenier, C. Fuchs, G. Jaeger, A. Y. Khrennikov, J.-\AA. Larsson, and S. Stenholm, AIP Conf. Proc. No.\ 1101 (AIP, New York, 2009), p.~241.


\bibitem{LSW11}
 Y.-C. Liang, R. W. Spekkens, and H. M. Wiseman,
 \href{http://dx.doi.org/10.1016/j.physrep.2011.05.001}{Phys. Rep. \textbf{506}, 1 (2011).}

\bibitem{AQBTC13}
 M. Ara\'ujo, M. T. Quintino, C. Budroni, M. Terra Cunha, and A. Cabello,
 \href{http://dx.doi.org/10.1103/PhysRevA.88.022118}{Phys. Rev. A \textbf{88}, 022118 (2013).}


\bibitem{SBBC13}
 M. Sadiq, P. Badzi\c{a}g, M. Bourennane, and A. Cabello,
 \href{http://dx.doi.org/10.1103/PhysRevA.87.012128}{Phys. Rev. A \textbf{87}, 012128 (2013).}

\bibitem{RDLTC14}
 R. Rabelo, C. Duarte, A. J. L\'opez-Tarrida, M. Terra Cunha, and A. Cabello,
 \href{http://dx.doi.org/10.1088/1751-8113/47/42/424021}{J. Phys. A \textbf{47}, 424021 (2014).}


\bibitem{Cabello13}
 A. Cabello,
 \href{http://dx.doi.org/10.1103/PhysRevLett.110.060402}{Phys. Rev. Lett. \textbf{110}, 060402 (2013).}


\bibitem{Yan13}
 B. Yan,
 \href{http://dx.doi.org/10.1103/PhysRevLett.110.260406}{Phys. Rev. Lett. \textbf{110}, 260406 (2013).}

\bibitem{ATC13}
 B. Amaral, M. Terra Cunha, and A. Cabello,
 \href{http://dx.doi.org/10.1103/PhysRevA.89.030101}{Phys. Rev. A \textbf{89}, 030101(R) (2014).}

\bibitem{Cabello15}
 A. Cabello,
 \href{http://dx.doi.org/10.1103/PhysRevLett.114.220402}{Phys. Rev. Lett. \textbf{114}, 220402 (2015).}


\bibitem{GKCLKZGR10}
 O. G{\"u}hne, M. Kleinmann, A. Cabello, J.-\AA. Larsson, G. Kirchmair, F. Z\"ahringer, R. Gerritsma, and C. F. Roos.
 \href{http://dx.doi.org/10.1103/PhysRevA.81.022121}{Phys. Rev. A \textbf{81}, 022121 (2010).}

\bibitem{LKGC11}
 J.-\AA. Larsson, M. Kleinmann, O. G{\"u}hne, and A. Cabello,
 in {\em Advances in Quantum Theory}, edited by
 G. Jaegger, A. Y. Khrennikov, M. Schlosshauer, and G. Weihs,
 AIP Conf. Proc. No.\ 1327 (AIP, New York, 2011), p. 401.

\bibitem{KZGKGCBR09}
 G. Kirchmair, F. Z\"ahringer, R. Gerritsma, M. Kleinmann, O. G\"uhne, A. Cabello, R. Blatt, and C. F. Roos.
 \href{http://dx.doi.org/10.1038/nature08172}{Nature (London) \textbf{460}, 494 (2009).}

\bibitem{CEGSXLC14}
 G. Ca\~{n}as, S. Etcheverry, E. S. G\'omez, C. E. Saavedra, G. B. Xavier, G. Lima, and A. Cabello,
 \href{http://dx.doi.org/10.1103/PhysRevA.90.012119}{Phys. Rev. A \textbf{90}, 012119 (2014).}

\bibitem{CAEGCXL14}
 G. Ca\~{n}as, M. Arias, S. Etcheverry, E. S. G\'omez, A. Cabello, G. B. Xavier, and G. Lima,
 \href{http://dx.doi.org/10.1103/PhysRevLett.113.090404}{Phys. Rev. Lett. \textbf{113}, 090404 (2014).}


\bibitem{Neves05}
 L. Neves, G. Lima, J. G. Aguirre G\'omez, C. H. Monken, C. Saavedra, and S. P\'adua,
 \href{http://dx.doi.org/10.1103/PhysRevLett.94.100501}{Phys. Rev. Lett. \textbf{94}, 100501 (2005).}

\bibitem{Lima09}
 G. Lima, A. Vargas, L. Neves, R. Guzm\'an, and C. Saavedra,
 \href{http://dx.doi.org/10.1364/OE.17.010688}{Opt. Express \textbf{17}, 10688 (2009).}

\bibitem{Lima10}
 G. Lima, E. S. G\'omez, A. Vargas, R. O. Vianna, and C. Saavedra,
 \href{http://dx.doi.org/10.1103/PhysRevA.82.012302 }{Phys. Rev. A \textbf{82}, 012302 (2010).}

\bibitem{Lima11}
 G. Lima, L. Neves, R. Guzm\'an, E. S. G\'omez, W. A. T. Nogueira, A. Delgado, A. Vargas, and C. Saavedra,
 \href{http://dx.doi.org/10.1364/OE.19.003542}{Opt. Express \textbf{19}, 3542 (2011).}

\bibitem{Lima13}
 S. Etcheverry, G. Ca\~nas,	E. S. G\'omez, W. A. T. Nogueira, C. Saavedra, G. B. Xavier, and G. Lima,
 \href{http://dx.doi.org/10.1038/srep02316}{Sci. Rep. \textbf{3}, 2316 (2013).}


\bibitem{Moli07}
 G. Molina-Terriza, J. P. Torres, and L. Torner,
 \href{http://dx.doi.org/10.1038/nphys607}{Nat. Phys. \textbf{3}, 305 (2007).}

\bibitem{Damb12}
 V. D'Ambrosio, E. Nagali, S. P. Walborn, L. Aolita, S. Slussarenko, L. Marrucci, and F. Sciarrino,
 \href{http://dx.doi.org/10.1038/ncomms1951}{Nat. Commun. \textbf{3}, 961 (2012).}

\bibitem{Naga12}
 E. Nagali, V. D'Ambrosio, F. Sciarrino, and A. Cabello,
 \href{http://dx.doi.org/10.1103/PhysRevLett.108.090501}{Phys. Rev. Lett. \textbf{108}, 090501 (2012).}

\bibitem{dimwit}
 V. D'Ambrosio, F. Bisesto, F. Sciarrino, J. F. Barra, G. Lima, and A. Cabello,
 \href{http://dx.doi.org/10.1103/PhysRevLett.112.140503}{Phys. Rev. Lett. \textbf{112}, 140503 (2014).}

\bibitem{micro}
 S. F\"urhapter, A. Jesacher, S. Bernet, and M. Ritsch-Marte,
 \href{http://dx.doi.org/10.1364/OPEX.13.000689}{Opt. Exp. \textbf{13}, 689 (2005).}

\bibitem{Tamb}
 F. Tamburini, B. Thid\'e, G. Molina-Terriza, and G. Anzolin,
 \href{http://dx.doi.org/10.1038/nphys1907}{Nat. Phys. \textbf{7}, 195 (2011).}

\bibitem{Damb13}
 V. D'Ambrosio, N. Spagnolo, L. Del Re, S. Slussarenko, Y. Li, L. C. Kwek, L. Marrucci, S. P. Walborn, L. Aolita, and F. Sciarrino,
 \href{http://dx.doi.org/10.1038/ncomms3432}{Nat. Commun. \textbf{4}, 2432 (2013).}

\bibitem{tweez}
 M. J. Padgett and R. Bowman,
 \href{http://dx.doi.org/10.1038/nphoton.2011.81}{Nat. Photon. \textbf{5}, 343 (2011).}

\bibitem{slm}
 R. Bowman, V. D'Ambrosio, E. Rubino, O. Jedrkiewicz, P. Di Trapani, and M. J. Padgett,
 \href{http://dx.doi.org/10.1140/epjst/e2011-01510-4}{Eur. Phys. J. Spec. Top. \textbf{199}, 149 (2011).}

\bibitem{qusix}
 V. D'Ambrosio,	F. Cardano,	E. Karimi, E. Nagali, E. Santamato, L. Marrucci, and F. Sciarrino,
 \href{http://dx.doi.org/10.1038/srep02726}{Sci. Rep. \textbf{3}, 2726 (2013).}

\bibitem{BBSKB13}
 E. Bolduc, N. Bent, E. Santamato, E. Karimi, and R. W. Boyd,
 \href{http://dx.doi.org/10.1364/OL.38.003546}{Opt. Lett. \textbf{38}, 3546 (2013).}


\bibitem{Winter14}
 A. Winter,
 \href{http://dx.doi.org/10.1088/1751-8113/47/42/424031}{J. Phys. A \textbf{47}, 424031 (2014).}

\bibitem{DKL15}
 J. V. Kujala, E. N. Dzhafarov, and J.-\AA. Larsson,
 \href{http://dx.doi.org/10.1103/PhysRevLett.115.150401}{Phys. Rev. Lett. \textbf{115}, 150401 (2015).}


\end{thebibliography}
\end{document}